\documentclass[preprint,12pt]{svjour3}

\usepackage{amssymb}
\usepackage{hyperref}

\begin{document}

\title{Mach's principle and the origin of the quantum phenomenon}

\author{Santanu Das}

\institute{IUCAA, 
Post Bag 4, Ganeshkhind, Pune 411 007, India \email{santanud@iucaa.ernet.in}}

\maketitle

\begin{abstract}
Concept of inertial mass is not well defined in physics.
For defining inertial mass of a particle we need to know its acceleration
under some force in an inertial reference frame, which itself is defined by 
the motion of its background objects. 
Therefore, the inertial
mass of a particle is not completely its intrinsic property and 
depends on the position of the particle itself and and all other particles (which we refer as background) in the universe. 
But the background of a particle keeps on fluctuating randomly due to different physical
phenomenon in the universe. Therefore, the exact position
or the mass of a particle can not be determined at any time, without
having full information about all the other particles
in the universe. Hence, in this paper, we define the dynamics of a particle
statistically. We show that the fluctuations in the background of an object 
contributes to an extra energy term in the total energy of the particle. 
If we treat this extra energy term as
the quantum potential then it leads to the Schrodinger equation.
Some examples are also given to show how a classical interpretation can be 
given to some quantum mechanical phenomenon.
\end{abstract}

\section{Introduction}	

The concept of mass is one of the most fundamental concepts in physics. 
Though the definition of mass is in use in physics from the days of Newton, 
there is no unique definition of mass.
However, for doing any physics 
we need some working concept of mass. 
The concepts of mass can be defined
in two different ways. The mass, defined from the inertial
 properties of matter, is called the inertial
mass. 
On the other hand, mass, defined from the gravitational
 properties of matter is called the gravitational mass. 
There can be two types of gravitational mass, namely active gravitational
mass and passive gravitational mass. 
A detailed discussion about different kinds of masses can be found in \cite{Jammer-1999}. 
Several high precession experiments are undertaken by physicists in the past century 
to check the equality between
inertial mass and passive gravitational mass and all these experiments
proved the equality up to a surprising accuracy. 
However, as the concept of
these two different masses evolves from two complete different parts
of physics and there is no direct logic to show the equivalence
between these two masses, the equality is still just an empirical phenomenon.

Inertial mass of a particle is 
measured based on its motion
of in an inertial coordinate frame.
Therefore, determining the inertial coordinate system is important 
for measuring the inertial mass. 
However, it is difficult to determine a perfect 
inertial coordinate system because there are no external reference frame based 
on which we can measure its acceleration. 
Ernest Mach postulated that the inertial reference could be determined
by measuring the motions of the distant objects in the universe. This implies that
the distant objects in the universe actually determine the inertial
properties of matter, which is the famous Mach's principle. Mach's principle
is discussed by many authors such as 
\cite{Sciama1953,Lichtenegger2004,Gilson2004,Barbour2010,brans1961,Raine1975,Neeman2005,Hoyle1964,Hoyle2000,Mashhoon2012,Wesson2011}
 in different contexts. 
Several, mathematical formalism are put forward for explaining it. But 
whatever be the formalism, if
we consider that the Mach's principle is correct and if there
are some fluctuations in the background (created by the distant objects) of a particle (which is always
present due to the motions of different objects in the universe) then
the inertial properties of the particle keep on changing and cannot 
be calculated deterministically without knowing the position of
all the other particles in the universe. 
A statistical description for the motion of the particles is required. 
In this paper we have postulated a new mathematical formalism of Mach's principle 
and shown that it under some assumption leads to Schrödinger's equation. Therefore, 
some of the quantum mechanical behavior may be originated due to the Mach's hypothesis.

There are several attempts to describe Schrödinger's equation \cite{Schrodinger1926}
from the classical picture. First of such attempts was made by David
Bohm in terms of hidden variable theory \cite{Bohm1952,Bohm1952a} where 
he introduced the concept of quantum potential responsible 
for producing quantum behavior in a particle. 
The formalism is popularly known as the Bohmian mechanics.
A similar theory known as pilot wave theory was also proposed by 
De Broglie. 
Another independent attempt was made by Nelson \cite{Nelson1966}
to explain Schrödinger's equation under the hypothesis that every
particle of mass $m$ is subject to a Brownian motion with diffusion
coefficient $\hbar/2m$ and no friction. Attempts were also made to
bring the Schrödinger's equation directly from the uncertainty principle \cite{Hall2002,Hall2002a}.
In some recent works, the Schrödinger's equation is derived
from the non-equilibrium thermodynamics \cite{Gerhard2008,Gerhard2009}.
An attempt to bring the Schrödinger's equation from Mach's principle
is also made by Gogberashvili \cite{Gogberashvil2011}. Gogberashvili's
work shows that the Machian model of the universe can provide a nice
platform for describing the quantum behavior of nature. There are many 
other interpretation of quantum mechanics such as \cite{Nottale1998,Nottale2007}. 
In this paper we make an attempt to explain some of the quantum mechanical behaviour from
Mach's principle using Bohmian formalism.

The paper is organized as follows. The second section describes the issues with the equality between 
inertial mass and passive gravitational mass. Concept of the background dependent inertial mass and its mathematical 
formulation is described in the third section. In the next section, we describe the 
dynamics of a particle statistically and derived the Schrödinger's equation using 
Bohmian formalism. The fifth section gives 
classical description to some quantum mechanical problems. Final section is the 
discussion and the conclusion section.

\section{Issues with the equality between Inertial and passive gravitational mass}

The equality between the passive gravitational mass and the inertial mass 
of a particle is given by the equivalence principle that comes from the logic 
that a particle with passive gravitational mass $m_{p}$
and inertial mass $m_{i}$, in a gravitational field of strength $g$, 
will have an acceleration $a=\left(\frac{m_{p}}{m_{i}}\right)g$. Therefore, 
any particle, at a given location
will fall (in vacuum) with same acceleration 
if and only if $m_{p}/m_{i}$ are same for all of them.  
If this is indeed the case, it will be convenient to choose an appropriate unit
to get $m_{p}=m_{i}$ and the result is confirmed by different experiments upto an 
extrimly high accuracy.
However, the experimental results does not tell anything if the ratio varies
over position and time, i.e. $\frac{m_{p}}{m_{i}}(t,x,y,z)$. Therefore, 
if it is considered that the ratio 
depends on the position of the particle in the universe depending on the background star galaxies etc.
then that will not violate any of the experimental results.

The concept can be understood as follows.
Suppose there is an observer inside a box and she is conducting
some experiment there, to check the acceleration of the box.
According to the Einstein's Equivalence Principle \cite{Einstein1954}, there is no experiment by which
she can distinguish whether her experimental results 
are due to the presence of some gravitating object outside
the box or due to the acceleration of the box, provided the experiment
is performed in a small enough region of space time. 
However, if there are some outside observer say $S$ 
and if there are two identical set up at two different parts of the universe say, $A$ and $B$,
where the backgrounds are not same, then the outside observer $S$ may see that the acceleration 
of $A$ and $B$ are not same. In this case the only possible explanation 
of the sitution for observer $S$ is the background. 

The explanations can be

\begin{enumerate}
\item The gravitational constant changes due to the background
\item The gravitational mass changes due to the background (either active
gravitational mass of the gravitating object or the passive gravitational
mass of the box)
\item The inertial mass of the box changes due to the background 
\end{enumerate}

The first option was analyzed in detail by Brans and Dicke \cite{brans1961}. 
If we consider that the gravitational mass of an object is its intrinsic property, the change of 
the gravitational mass is not possible.
Therefore, we explore the last possibility, as there are strong logic behind the fact that 
the inertial mass of an object is not completely its intrinsic property 
and depends on the other masses of the universe.

\section{Inertial mass and its dependence on the background objects}

The inertial mass was defined by Mach in the following way. Any
two bodies $A$ and $B$, in this universe 
apply force on each other, which induces opposite accelerations
on both the bodies along their line of junction. Provided we 
can define an inertial reference frame while looking from it 
if $a_{A/B}$ be the acceleration of the body $A$
towards $B$ and $a_{B/A}$ be the opposite, then according to the
Mach's definition of inertial mass, the quantity $-\left(\frac{a_{A/B}}{a_{B/A}}\right)$,
which is a positive quantity, which gives the ratio of the inertial
masses of the two bodies, i.e. $m_{A/B}=\frac{m_{A}}{m_{B}}$.

If a third body $C$, which is interacting with $A$
and $B$, is introduced then 
the mass ratios satisfy transitive relations,
i.e. $m_{A/B}=m_{A/C}m_{C/B}$. Therefore, the mass
ratio of two bodies remains a positive quantity. If
we take one body as the standard, then we can define the mass
of the other bodies \cite{Jammer-1999}. 

However, this definition of inertial mass is not valid for a system
consisting of $n>4$ bodies (\cite{Jammer-1999,Pendse1937}). 
If a system has $n$ bodies, the observable induced acceleration
of the $k^{th}$ body is given by $a_{k}$ and $u_{ik}$ is
the unit vector from the $i^{th}$ body to the $k^{th}$ body, then 
the accelerations of the bodies in a 3 dimensional space can be written as

\begin{equation}
a_{k}=\sum_{j=1}^{n}\alpha_{kj}u_{kj}\,\,\,\,\,\,\,\,\,\,\,\,\,\,\,\,\,\,\,\,\,\,\,\,\:......k=1(1)3n\,.\label{eq:pendse}
\end{equation}

\noindent Here, $\alpha_{ik}$ ($\alpha_{ik}\neq0$) are $n(n-1)$ unknown numerical
coefficients for $3n$ equations that can
be used for determining the mass ratios. However, these numerical coefficients
are uniquely determined only if number of unknowns does not exceed
the number of equations i.e. $n(n-1)\leq3n$, or $n\leq4$. Therefore,
in a universe with more than $4$ particles 
the inertial masses of the particles can be determined uniquely.
The mass of any particle will depend on mass and position of other particles, i.e. the background. 

The Eq.(\ref{eq:pendse}) has infinitely many solutions and only a 
particular solution will provide the real mass of the particles. However,
its impossible to find out, which of these solutions will provide
the real mass of the particles at any time instant.

The above problem can be solved if we consider that initially all the particles 
in the universe had some masses and 
whenever some particle is moving through space-time, 
it sends some signal 
through space-time, which carries information about the mass.
When the signal reaches other masses, they change their
inertial masses accordingly. In this way each  mass can
be connected to all the other masses in the universe. Hence, each
 mass will have some definite value of the inertial mass and
the above problem gets sorted out. This will make the inertial
masses deterministic and still dependent on the background.

If it is considered that the universe is almost homogeneous
and isotropic and several small-scale random fluctuations keep on
occurring in different parts of the universe, then the inertial
mass or the inertial properties of an object will keep on changing
and only a probabilistic behavior of the position
or the momentum of the particles can be calculated. 

\subsection{Mach's principle and the energy equation}

Suppose a particle is located at a place A, in the universe.
Assume that all the particles in the background are far away to show
any type of gravitational attraction.
According to the previous discussion, all the particles in the background
have some contribution to the mass of the particle kept at $A$. Suppose
in this configuration the inertial mass of the particle is $m_{1}$.
If all the particles in the background 
move apart to give the background a new configuration then
in that new configuration of the background, the inertial
mass of the particle at $A$ will change. 
Let this new mass is $m_{2}$. 

If it is considered that the total energy and the momentum of the particle for these
two configurations of the background are $E_1$, $E_2$ and $p_1$, $p_2$ respectively then

\begin{equation}
E_{i}^{2}=m_{i}^{2}c^{4}+p_{i}^{2}c^{2}\,,\label{eq:energy-con-1}
\end{equation}

\noindent where $i=1,2$. 

But as mass of the particle is changing 
we cannot define the 
mass in this case as function of only $(t,x,y,z)$. There has to be one extra 
variable which will come from the background because in this case the $(t,x,y,z)$
are constant for the particle but still mass is changing. Therefore, we define the mass
as $m(t,x,y,z,\zeta)$, where $\zeta$ is the new variable showing the contribution 
of the background. 

As we introduce this new coordinate dimension $\zeta$, in the energy equation we must 
introduce the moment which comes from the variation of 
this new coordinate $\zeta$ with respect to the line element $ds$. Also as 
we have an extra variable ($\zeta$)
that takes care of the variation of the background, we can 
absorb the variation of the mass in the definition of the coordinate $\zeta$ and redefine 
the Energy and momentum as $E \rightarrow \frac{mE}{m(t,x,y,z,\zeta)}$ and 
$p \rightarrow \frac{mp}{m(t,x,y,z,\zeta)}$ where $m$ is a constant mass. 
Thus the energy equation becomes 

\begin{equation}
E^{2}=m^{2}c^{4}+p^{2}c^{2}+\epsilon E_{m}^{2}\,.\label{eq:energy-con}
\end{equation}

\noindent Here $E_{m}$ can be thought of as the momentum coming from the variation of $\zeta$
and is the only background dependent term. 
Here, $\epsilon$ will take either $+1$ or $-1$ depending on whether the coordinate 
is time like or space-like. The signature will be fixed in a later section.

 Eq.($\ref{eq:energy-con}$) leads to

\begin{equation}
c^{2}\frac{dt^{2}}{ds^{2}}=1+\frac{dx^{2}}{ds^{2}}+\frac{dy^{2}}{ds^{2}}+\frac{dz^{2}}{ds^{2}}+\epsilon K^{2}\frac{d\zeta^{2}}{ds^{2}}\,.\label{eq:line_element_1}
\end{equation}

\noindent Here, $\zeta$ is the new dimension which is measuring the effect from
the background and $K^{2}\frac{d\zeta^{2}}{ds^{2}}=\frac{E_{m}^{2}}{m^{2}c^{4}}$.
$\zeta$ may not have the dimension of space. Therefore, the constant $K$ is multiplied to
make the equations dimensionally correct.

Eq.(\ref{eq:line_element_1}) gives the line element as 

\begin{equation}
ds^{2}=c^{2}dt^{2}-dx^{2}-dy^{2}-dz^{2}-\epsilon K^{2}d\zeta^{2} = ds_{c}^{2}-\epsilon K^{2}d\zeta^{2}\,\,,
\end{equation}

\noindent where $ds_{c}$ is the special relativistic line element.

A similar attempt to explain Mach's principle by using 5 dimension coordinate system 
was also recently made by \cite{Mashhoon2012,Wesson2011}. Though their method 
is different from us both logically and mathematically.

\subsection{Fixing the signature of the 5th dimension}

Suppose a particle moves along the $x$ axis with respect to some observer. 
Also consider that the particle has some momentum along the $\zeta$ axis 
as it is controlled by the background objects. So, the line element is given by 

\begin{equation}
d\tau^{2}=dt^{2}-\frac{dx^{2}}{c^{2}}-\frac{K^{2}}{c^{2}} \epsilon d\zeta^{2}\,,
\end{equation}

\noindent where, $\tau$ is the proper time of the particle.

The coordinate time
and the proper time of a particle can be related by the expression

\begin{equation}
dt=\frac{d\tau}{\sqrt{\left(1-\epsilon\frac{K^{2}}{c^{2}}\frac{d\zeta^{2}}{dt^{2}}-\frac{1}{c^{2}}\frac{dx^{2}}{dt^{2}}\right)}}\,.
\end{equation}

\noindent Provided $\epsilon=+1$, the velocity of a particle in a 5 dimensional
system can't reach $c$ at any stage unless $\frac{d\zeta}{dt}=0$. If $\epsilon=-1$, the particle 
can have a velocity higher than $c$ and still $dt$ can be real.
Therefore, to avoid any such problem with superluminal velocity of the particle, we fix $\epsilon$ to $+1$.

\subsection{Calculating the action}

As the space is considered to be 5 dimensional, the five dimensional line element 
i.e. $ds$ remains constant in case of any coordinate transformation. 
Therefore, for a flat manifold the action can be defined as 

\begin{equation}
S=-mc\int ds\,,
\end{equation}

\noindent where $ds$ is the line element and $m$ is the mass of the object, a constant. 

In the non-relativistic
limit, the standard classical mechanics line element $ds_{c}$ can be written as 

\begin{equation}
ds_{c}=cdt(1-\frac{1}{2}\frac{v^{2}}{c^{2}})\,.
\end{equation}

\noindent However, if the background contribution is taken into account for the calculations
then the new line element becomes

\begin{equation}
ds=cdt(1-\frac{1}{2}\frac{v^{2}}{c^{2}}-\frac{1}{2}\frac{K^{2}}{c^{2}}\frac{d\zeta^{2}}{dt^{2}})\,.\label{eq:line_element}
\end{equation}

\noindent Here we consider that 
$\frac{K^{2}}{c^{2}}\frac{d\zeta^{2}}{dt^{2}} \ll 1$. 
However, if the term is large enough then the above approximation may not hold.

According to the line element given in Eq.(\ref{eq:line_element}), the
full action of the particle will be the action from the standard classical mechanics plus
a small quantity

\begin{equation}
S=S_{c}+\delta S\,,
\end{equation}

\noindent where $\delta S=\frac{1}{2}m K^{2}\int dt\left(\frac{d\zeta}{dt}\right)^{2}$. 

Any calculation of the equation
of motion from standard classical mechanics, can not provide
total  momentum or energy
of a particle because there is one more dimension yet to be fixed.
Suppose a particle is going from point $A$ to point
$B$. By minimizing the action, the path of the particle 
from one place to another can be found out. If it is considered that
the background dimension fluctuates randomly due to random events
in the universe, then if a particle goes from $A$ to $B$ for $n$ 
times then it will go through $n$ different paths. 
This leads to the quantum mechanical behavior in its motion.

\section{Statistics of the particle motion}

Due to the expansion of the universe, the background of all particles changes 
with time  which will have some overall effect on the inertia. 
However, in this paper instead of considering such long term change, we 
focus on small fluctuations in the background. It is known that the universe 
is almost homogeneous and isotropic. But there are small time dependent fluctuations 
due to the motions of different objects such as stars galaxies etc. So due to those fluctuations in the background,
the four dimensional space time hyperspace shows random variations. Though the five dimensional 
space time background manifold will not vary as otherwise it will have non-vanishing Ricchi scalar, 
which is not the case as we are not considering any gravitational mass. 
Due to these random fluctuations in the four dimensional hyperspace, 
the momentum and the energy of the particle fluctuates around some average value. 
Therefore, to get the properties of the energy and momentum of a particle 
we need to calculate the statistics of this variation.

\subsection{Calculating the probability density function}

In the five dimensional flat coordinate system 
a particle should follow the equation $m^2=E^2-p^2_x-p^2_y-p^2_z-p^2_{\zeta}$, where $m$ is a constant.
Here $\zeta$ direction is defined as the direction perpendicular to the local space-time
hyperspace. As in this paper we are only interested
in the non-relativistic limits,  we can write the above equation 
as $E = \frac{1}{2m}(p^2_x+p^2_y+p^2_z+p^2_{\zeta})$.

In standard classical mechanics as
we analyze the 4-D motion of a free particle
and ignore the  contribution from the background dimension, 
the energy from the background term i.e. $p_{\zeta}$ 
acts as a space-time dependent potential energy in the energy equation.
Conceptually it can be explained as follows. Suppose a ball is moving
in a `rolling ball sculpture'. If we track the projection of the ball on a wall 
then  it moves slowly when the ball moves faster in
the perpendicular direction to the wall.
Therefore, if someone wants to define the laws
of motion of the particle just by looking its motion on the two-dimensional
wall then the energy from the motion of the particle along the perpendicular 
dimension of the wall behaves as
a potential energy acting on the projection. As in our theory, the
particle is actually moving in a five dimensional space-time-background,
when we analyze the motion of the particle from the four-dimensional
space-time we need to add extra energy that is
equal to the kinetic energy of the particle from the background
dimension, in the Hamiltonian.

Suppose a particle at time $t$ was at $A$$=$$(x,y,z,\zeta)$
and its momentum along $\zeta$ direction was $m\frac{d\zeta}{dt}$.
So when we do our calculations in four dimension, this quantity
behaves as a field which can provide extra momentum along different directions.
So if the particle moves to $B$$=$$(x+dx,y+dy,z+dz,\zeta+d\zeta)$
after some time $dt$ through the path $s$, then the line integral $I_{AB}=\int_{A}^{B}m\Big|\frac{d\zeta}{dt}\Big|ds_c$
can be related to the probability of finding the particle at $B$,
i.e. $P_{AB}$ at $t+dt$, given that the particle was at $A$$=$$(x,y,z,\zeta)$
at time $t$. Suppose the particle goes from $A$ to $C$ via
$B$. Then the line integral goes as summation i.e. $I_{ABC}=\int_{A}^{B}m\Big|\frac{d\zeta}{dt}\Big|ds_c
+\int_{B}^{C}m\Big|\frac{d\zeta}{dt}\Big|ds_c$$=I_{AB}+I_{BC}$.
However, the probability will go as multiplication
i.e. $P_{ABC}=P_{AB}P_{BC}$. Therefore the best way to relate these
two quantities is $I_{AB}=-\log P_{AB}$. We can always absorb all the
multiplication constants in the definition of $\zeta$. This leads to 

\begin{equation}
m\Big|\frac{d\zeta}{dt}\Big|=\Big|\frac{\nabla P}{P}\Big|. \label{eq:assumprion}
\end{equation}

\noindent Here, $P$ is the probability that the particle exists at $(x,y,z)$ at time $t$.   
The negative sign is taken because the left hand side is always positive as it is 
the integral of some absolute quantity but in the right hand side $P_{AB}$ must be less than 
unity and thus $\log P_{AB}$ must be negative. Therefore, a negative sign is required to 
match both the sides.

Here the modulus is taken because $\nabla P$ is a vector quantity. 
To know the direction of $\nabla P$
we need to know the actual shape of the geodesic at a particular point, which
cannot be known without knowing the position and the states of all
the particles in the universe at that instant. Eq.(\ref{eq:assumprion}) is a pure assumption of this paper. 
Using this assumption the Hamiltonion can be rewritten as

\begin{equation}
H=\frac{1}{2m}\left(p_{x}^{2}+p_{y}^{2}+p_{z}^{2}+\epsilon K^{2}\left(\frac{\nabla P}{P}\right)^{2}\right)+V\,. \label{eq:ham}
\end{equation}

\subsection{Calculating the wave function}

A Bohm like interpretation of quantum mechanics can be used here for deriving the 
quantum mechanical equations. Eq.(\ref{eq:ham}) shows that here
the Bohemian quantum mechanical potential comes from 
the background term.

If the background fluctuations were not present then the Hamiltonian for a particle moving in a potential $V(\vec{x})$, 
could have been written as the standard classical Hamiltonian, i.e.

\begin{equation}
H=\frac{p_{x}^{2}}{2m}+\frac{p_{y}^{2}}{2m}+\frac{p_{z}^{2}}{2m}+V(\vec{x})\,.
\end{equation}

\noindent  In that case we could have taken $\frac{d\zeta}{dt}$ as $0$ and hence $P=1$, i.e. no quantum nature of the particle.
This Hamiltonian of Eq(\ref{eq:ham}) is a random Hamiltonian, because $\left|\frac{\nabla P}{P}\right| = m\left|\frac{d\zeta}{dt}\right|$,
is a random variable, with a given probability distribution. 
As the probability density function of the Hamiltonian to occur
for a fixed classical action is determined, the expectation value
of the Hamiltonian can be calculated as 

\begin{eqnarray}
\bar{H} & = & \int P(\vec{x})\left(\frac{p_{x}^{2}}{2m}+\frac{p_{y}^{2}}{2m}+\frac{p_{z}^{2}}{2m}+V(\vec{x})+\frac{K^{2}}{2m}\left(\frac{\nabla P}{P}\right)^{2}\right)dx^{3} \nonumber\\
  & = & \int P(\vec{x})\left(\frac{\left(\nabla S_{c}\right)^{2}}{2m}+V(\vec{x})+\frac{K^{2}}{2m}\left(\frac{\nabla P}{P}\right)^{2}\right)dx^{3}\,,\label{eq:Expected_Hamiltonian}
\end{eqnarray}

\noindent where $S_{c}$ is the standard classical mechanics action from the 4 dimensional theory
as discussed in previous section and $V(x)$ is the potential.

For a moment we can treat $P$ as some field and $S_{c}$ to be
its conjugate momentum \cite{Bohm1952,Bohm1952a}. Such an assumption is a valid assumption and
that can be seen from the equations of motion derived from this. Hamiltonian mechanics gives

\begin{eqnarray}
\dot{P} & = & \frac{\delta\bar{H}}{\delta S_{c}} =  -\frac{1}{m}\nabla(P\nabla S_{c})\,.
\label{eq:contineuty}
\end{eqnarray}

\noindent The expression is obtained by taking `integration by parts' of Eq.(\ref{eq:Expected_Hamiltonian}) and then differentiating with respect to $S_{c}$.
Similarly, the time derivative of $S_{c}$ can be obtained as

\begin{eqnarray}
\dot{S_{c}} & = & -\frac{\delta\bar{H}}{\delta P}
  =  -\left[\frac{1}{2m}\left(\nabla S_{c}\right)^{2}+V(x)-\frac{K^{2}}{m}\left(\frac{\nabla^{2}P}{P}-\frac{1}{2}\left(\frac{\nabla P}{P}\right)^{2}\right)\right]\,.
\label{eq:momentum}
\end{eqnarray}

These equations
can also be derived from the conservation equation but the above way of
deriving these equations is easier. We can physically interpret the first equation i.e. Eq.(\ref{eq:contineuty}) as the conservation
equation and the second equation i.e. Eq.(\ref{eq:momentum}) as the Hamilton Jacobi equation.
Intuitively, these equations are same as the equations for the irrotational
bariotropic flow with density $P$ and the internal energy
 $P\left(V(\vec{x})+\frac{K^{2}}{2m}\left(\left(\frac{\nabla P}{P}\right)^{2}\right)\right)$.
Under such assumption, the above equations will match with the equations
of motions of fluid dynamics equations for a irrotational barotropic
flow. 

The above equations i.e. Eq.(\ref{eq:contineuty}) and Eq.(\ref{eq:momentum})
can be rearranged and combined together to give a single equation
of the form

\begin{equation}
i\frac{\partial\Psi}{\partial t}=-\frac{2K^{2}}{m}\nabla^{2}\Psi+V\Psi\,.\label{eq:Schrodinger_eqn}
\end{equation}

\noindent where, $\Psi=P^{\frac{1}{2}}\exp\left(iS_{c}/2K\right)$. 
The Eq.(\ref{eq:Schrodinger_eqn}) is nothing but the Schrodinger
equation provided we take the constant $K=\frac{\hbar}{2}$. 
Therefore, the Schrödinger's wave equation
is derived in a completely classical way from the Mach's principle 
using Bohemian formalism.

In case, the logic discussed in the previous section for adding the
extra potential equivalent to the kinetic energy in the $\zeta$ direction
is not convincing, the concept can be rethought as
follows. Suppose there is an ensemble of particle trajectories following
the equation of motions Eq.(\ref{eq:contineuty}) and Eq.(\ref{eq:momentum}). If we consider a completely
standard classical picture then the kinetic energy in the $\zeta$
dimension will vanish, i.e. the extra potential term in the Hamiltonian
(Eq.(\ref{eq:ham})) will vanish. This can be done by putting $\hbar=0$ in the above
equations. The trajectories behave as the standard classical
mechanics and hence they will be normal to any constant $S_{c}$ hyperspace
and at any point $(x,y,z)$ of that hyperspace, $\frac{\nabla S_{c}}{m}$
will give the velocity vectors of the particles. The Eq.(\ref{eq:contineuty})
gives the standard continuity  equation. Therefore, Schrodinger equation
in the standard classical limit approximation is just a composition
of the two standard classical equations. 

This interpretation  can be extended further. When we are looking
at the dynamics of the particles, the particles are not actually
moving through the four dimensional space-time but in a five dimensional
space-time-background. But when we calculate the dynamics of a particle,
the motion in the background dimension is not considered. Therefore,
the kinetic energy of the particle along the background dimension
projects itself as a quantum mechanical potential acting on each
particle along with the classical potential $V$. The trajectories
of all the particles can not be calculated independently without knowing
the details of background. But as the energy from the background dimension follows a 
probability distribution, the dynamics 
of the group of particles has to be computed statistically. This provides a new
interpretation  to the Schrödinger's equation.

\section{Example }

In this section we show how Machian 5D model can be used 
for providing classical interpretation to some quantum mechanical problems.

\subsection{Tunneling through a potential wall}

Tunneling phenomenon is one of the many quantum mechanical problems that
have no classical explanation. Here we 
describe how this phenomenon can be explained from classical point of view. 
Suppose we have a step potential, where potential is $0$ for negative $x$ and
 $+V_{0}$ for positive $x$. 
According to the quantum mechanics, when the particle
travels along the positive $x$ direction, the wave function
takes the form 

\begin{equation}
\Psi(x)=\exp(-\frac{\sqrt{2m(V_{0}-E_{c})}}{\hbar}x)\,,
\end{equation}

\noindent where, $E_{c}$ is the total energy of the particle calculated from
the 4 dimensional motion of the particle and differs from the actual
energy of the particle by the energy contribution from the $5^{th}$
dimension. 

According to quantum mechanics, the kinetic energy of the particle,
when it is at a distance $x$ from the origin is given by 

\begin{eqnarray}
K.E. = -\frac{1}{\Psi(x)}\frac{\hbar^{2}}{2m}\frac{\partial^{2}}{\partial x^{2}}\Psi(x)
  =  -(V_{0}-E_{c})\,.
\end{eqnarray}

Therefore, we have to consider that the kinetic energy of the particle 
is getting negative. However, if the background dimension gives some
contribution to the kinetic energy then the kinetic energy can be made 
positive and everything will behave classically. 

The wave function shows that the probability of finding the particle
at $x$ is given by 

\begin{eqnarray}
P = \Psi^{*}(x)\Psi(x) = \exp(-\frac{2\sqrt{2m(V_{0}-E_{c})}}{\hbar}x)\,.
\end{eqnarray}

\noindent Hence we will have 
\begin{equation}
\frac{\nabla P}{P}=-\frac{2\sqrt{2m(V_{0}-E_{c})}}{\hbar}\,.
\end{equation}

\noindent Applying the relation between $\frac{\nabla P}{P}$ and $m\frac{d\zeta}{dt}$ we get 

\begin{equation}
\frac{1}{2}m\left(\frac{d\zeta}{dt}\right)^{2}=(V_{0}-E_{c})\frac{4}{\hbar^2}\,. \label{eq:total-energy}
\end{equation}

\noindent According to the previous discussion, the total 
energy of a particle can be written as 

\begin{equation}
E=\frac{1}{2}\frac{p^{2}}{m}+V_{0}+\frac{1}{2}m\frac{\hbar^2}{4}\left(\frac{d\zeta}{dt}\right)^{2}\,.
\label{eq:total-energy_5D}
\end{equation}

\noindent Therefore, there is no need to consider that the total kinetic energy
is getting negative or the particle is having some imaginary momentum.

An alternate explanation to the phenomenon can be that
the particle will stop when the kinetic energy becomes equal to the potential energy.
If the particle stops at a distance $x$ then at $x$ , $E=V_{0}$.
According to 4 dimensional model total energy of the particle is 

\begin{equation}
E_{c}=\frac{1}{2}\frac{p^{2}}{m}+V_{0}\,.
\label{eq:total-energy_4D}
\end{equation}

\noindent Eq.(\ref{eq:total-energy_5D}) and Eq.(\ref{eq:total-energy_4D}) leads to Eq.(\ref{eq:total-energy}), and doing the calculation
backward we can find the probability of a particle that will stop at a distance $x$ is 

\begin{equation}
P=\exp(-\frac{2\sqrt{2m(V_{0}-E_{c})}}{\hbar}x)\,.
\end{equation}

\noindent Therefore, it gives a classical mechanics interpretation to tunneling phenomenon.

\subsection{Stationary states}

Stationary states are the states where the probability of finding
a particle does not change with time. Stationary states only occur when
the particle is bounded inside some infinite potential wall. The conditions
for the stationary states are 
\begin{enumerate}
\item The particle's energy is a constant of motion and hence independent of time. 
\item The probability density of finding the particle at some point 
remains stationary, i.e. $\dot{P}=0$.
\end{enumerate}
Therefore, the equations for finding the stationary states of a particle
have to be 

\begin{eqnarray}
0 & = & \nabla(P\nabla S_{c})\label{eq:contineuty-1}\,,\\
E & = & \frac{1}{2m}\left(\nabla S_{c}\right)^{2}+V(x)-\frac{\hbar^{2}}{4m}\left(\frac{\nabla^{2}P}{P}-\frac{1}{2}\left(\frac{\nabla P}{P}\right)^{2}\right)\,.\label{eq:momentum-1}
\end{eqnarray}

To solve the equations we need the boundary conditions, which
can be found as follows. Suppose, there is a infinite potential energy
boundary at a position $x = 0$. 
Let us consider a point $Q$ at a distance $\delta$, which is very
close to the boundary. If the particle has to move to that point then
it needs an $\infty$ energy contribution from the background,
and hence it needs $\frac{d\zeta}{dt}$ to be infinity at that place. As $\nabla\ln P = -m\frac{d\zeta}{dt} = -\infty$,
we should get $P = 0$ inside the potential wall. Considering the probability
distribution of a finding a particle at some point in space to be
continuous we can have $P(x=0)=0$.

As the wave function $\Psi$ of a particle is related with the probability
$P$ with the equation $\Psi=P^{1/2}\exp(iS_{c}/\hbar)$, 
$\Psi$ will also become $0$ at position $x=0$. 

The above equations  can be solved using the boundary conditions to 
get all the energy levels of a particle, though the calculations
are done in the framework of classical mechanics.

\section{Conclusion}

Mach's principle is studied in detail in this paper and a new
mathematical formalism is derived. It is shown
that the inertial properties of a particle depends on the background
of the position of the particle. Random fluctuation of the background 
caused by the motion of different objects in the universe, can give rise to 
the random changes in different inertial properties of a particle and can 
introduce quantum mechanical behavior in motion of a particle. 
Therefore, some of the quantum mechanical
behaviors of a particle can be given a classical description.

One of the measure objections to most of the hidden variable 
theories that try to explain the 
quantum mechanics is the quantum entanglement phenomenon. 
In our theory we use a 
Bohemian formalism for explaining the quantum mechanical behaviors of a particle. 
Its well known that the Bohemian formalism of quantum mechanics 
can explain the quantum entanglement phenomenon \cite{Bohm1952,Bohm1952a}. Apart from that
there are also other explanations to the entanglement phenomenon such 
as the direct particle approach of Hoyle and Narlikar \cite{Narlikar1969,Narlikar1971}.
Therefore, the quantum entanglement does not disproves the theory presented in this paper. 

The theory presented in this paper not only explains some of the quantum mechanical phenomenon, 
it shows
that even in the quantum level there exist some definable variables that are controlled 
by the motion of different other particles of the universe and introduces 
quantum behavior to the system. Hence, it relates 
the Mach's principle with quantum mechanics.
Therefore quantum mechanics does not behave in a merely probabilistic sense.
This provides quantum mechanics a sense of completeness.

\section*{Acknowledgment}
I would like to thank Krishnamohan Parattu, Rizwan Ansari, Pallavi Bhat, 
Prof. Tarun Souradeep and Prof. J.V. Narlikar for 
careful reading of the draft and for several interesting discussions.

\end{document}